# Basic Elements of Strong Gravitational Lensing

Paul L. Schechter[1,2*] and Jeremy D. Schnittman[3,2]

[1*]Department of Physics, Massachusetts Institute of Technology,
77 Massachusetts Avenue, Cambridge, 02139, MA, USA.
[2]MIT Kavli Institute, Massachusetts Institute of Technology,
77 Massachusetts Avenue, Cambridge, 02139, MA, USA.
[3]Gravitational Astrophysics Lab, NASA Goddard Space Flight Center,
Greenbelt, 20771, MD, USA.

*Corresponding author(s). E-mail(s): schech@mit.edu;
Contributing authors: jeremy.d.schnittman@nasa.gov;

**Abstract**

Even when used to describe the same phenomenon, equations, graphics and words each give different perspectives and lead to complementary insights. The basic elements of strong gravitational lensing are introduced here favoring words and graphics over equations whenever possible. Fermat's principle is the fundamental driver of strong lensing. Three "D's" encapsulate the essential effects of lensing: **D**elay, **D**eflection and **D**istortion. Gravity and geometry both contribute to the delay of photons from a lensed source. Their interplay determines how the images of a source are deflected and how they are stretched or compressed. Caustics and critical curves are explained. Images of doubly, triply, quadruply and quintuply lensed sources are displayed. A table of symbols, their definitions and distinctions provides a summary of the basic elements of strong lensing.

# 1 What is strong gravitational lensing?

Strong lensing occurs when the gravitational field of an astronomical object is sufficiently strong to produce two or more images of a single background source. The French words "mirage gravitationelle" describe the phenomenon more accurately. The multiple images are distorted; sometimes their handedness is flipped; and as projected onto the sky they are displaced from the actual position of the lensed source. The phenomenon is illustrated schematically in Figure 1.

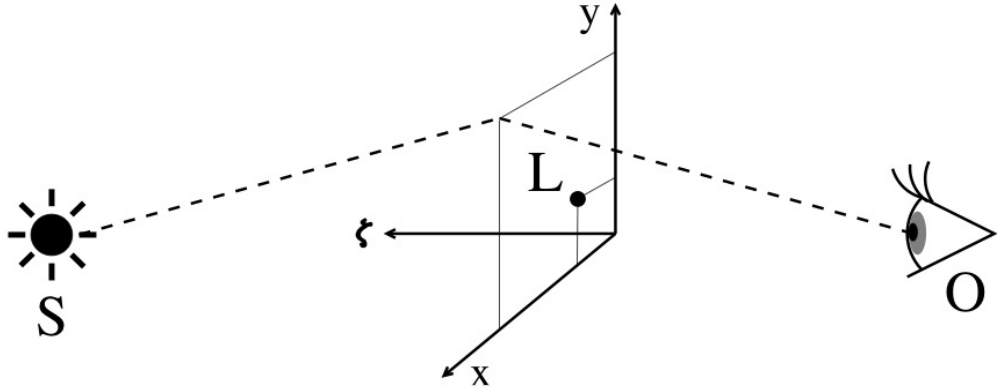

**Fig. 1** A gravitational lens $L$ deflects light from a source $S$. The $\zeta$ axis lies along the unperturbed line of sight from the source to the observer O. The lens lies at $\zeta = 0$.

Strong lensing has several basic elements that combine in myriad ways to produce complicated strongly lensed systems. These permit the exploration of a number of outstanding questions in astronomy. Our purpose in this chapter is to equip the reader with the tools needed to carry out such explorations.

Among the many idealizations that we adopt here for illustative purposes is the assumption that our strong lenses are single objects (as with the case of the elliptical galaxy in Figure 2). The treatment of "lenses within lenses" is addressed by Natarajan et al. (2024) who discuss lensing by clusters of galaxies and by Vernardos et al. (2024) who discuss the micro-lensing of quasars by stars.

## 2 An observational archetype for strongly lensed system

In Figure 2a we see a blue galaxy at redshift 0.197 lensed by a yellowish foreground galaxy with redshift 0.120 (Bolton et al. 2008). This system, SDSS J0044+0113 is among the most ordinary-looking strongly lensed systems known, and we adopt it as our observatinal archetype. The apparent graininess results from statistical noise in the original Hubble Space Telescope (henceforth HST) monochromatic exposure. Figure 2b shows a smooth, theoretical model for this system. The data and model have been "colorized" to conform with our expectations for galaxies at these redshifts.

Figure 2c shows a highly idealized circularly symmetric model drawn from Narayan and Bartelmann (1996) that illustrates some basic elements of strong lensing. The idealized source is shaped like a sector of an annulus, located an angular distance $\beta$ from the center of the lens.

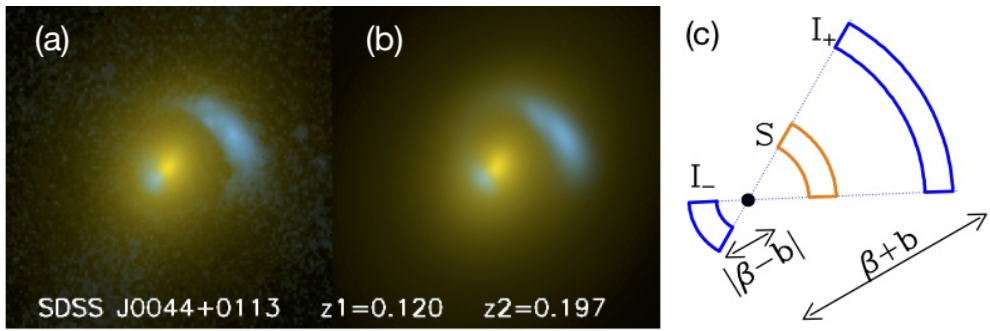

**Fig. 2** An archetypical gravitationally lensed system. Panel (a) shows a yellow galaxy lensing a blue galaxy that lies behind it. Panel (b) shows a model for the system. Panel (c) shows a highly idealized schematic of the model. The positive and negative parity images, $I_+$ and $I_-$, lie at angular distances $\beta + b$ and $|\beta - b|$ from the center of the lens, where $\beta$ is the offset of the actual, unlensed, position of the source, $S$. The strength of the lens model (a singular isothermal spehre) is given by the radius of its Einstein ring, $b$, which is greater than $\beta$.

The idealized lens (known as a **S**ingular **I**sothermal **S**phere; henceforth SIS) has the property that all rays are deflected radially by the same angle, $b$. One image, $I_+$, lies at angle $\beta + b$ from the lens and is magnified by a factor $1 + \frac{b}{\beta}$. Another image, $I_-$, lies at angular distance $|b - \beta|$ from the lens and is (de)magnified by a factor $1 - \frac{b}{\beta}$. Note $I_+$ has the same handedness as the source (positive image parity), while $I_-$ has the opposite handedness (negative image parity, as indicated by the sign of the magnification).

## 3 Strong lensing from Fermat's principle

Terrestrial mirages are the result of variations with height in the index of refraction $n$ of the Earth's atmosphere. The speed of a photon traveling through a medium with index $n$ is $c/n$. Fermat's principle tells us that images appear along directions for which a photon's time of flight is a minimum (or a maximum, or a saddle point). Those directions will, in general, be different from the direction along which a source would be seen in the absence of a refractive medium. These deflections from the unlensed light paths increase the path length and consequently delay the arrival of photons. It is therefore the *sum* of the refractive delay and the second, geometric delay, that is subject to Fermat's principle.

The gravitational potentials of astronomical objects (galaxies for the present discussion) create "effective" indices of refraction, which then delay the times of flight of rays passing through them (Binney and Merrifield 1998). Images of a background source appear along directions for which the time of flight is a stationary point.

Something akin to Fermat's principle is at work when one uses GPS enabled software to choose the shortest route from one city to another, as illustrated in Figure 3. The traffic on the roads acts like an index of refraction, delaying the travel. The software chooses the route that takes the least time.

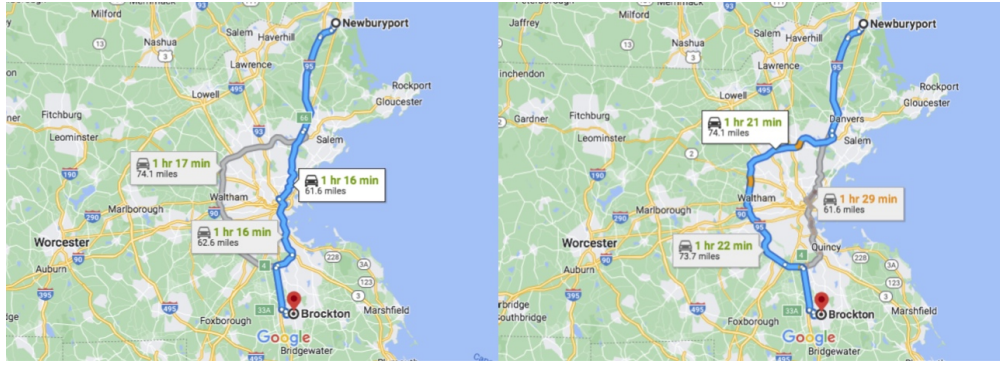

**Fig. 3** Automobile traffic plays the same role as an index of refraction, reducing speed and increasing the travel time. In the absence of traffic, the fastest route from Newburyport to Brockton goes through downtown Boston, which lies on a straight line between the two. Panel a) shows that route, chosen by a GPS application. In panel b) traffic is heavy, introducing an effective index of refraction. Though the Fermat route has a longer path length, the total travel time is shorter because traffic is lighter.

In the example shown, in the absence of traffic, the quickest route passes through the city of Boston. But during rush hour it takes less time if one bypasses the city. Note that Boston is not small compared to the distance traveled, so one cannot use the "thin lens" approximation described below.

## 4 The thin lens approximation and the time delay function

The scale on which a galaxy produces a non-trivial effective index of refraction is many orders of magnitude smaller than the distance between the lens and the source or the observer. This permits use of the "thin lens" approximation of conventional optics, greatly simplifying modelling. It is implicit in Figure 1, in which the direction of travel of photons changes discontinuously in the plane of lens.

The essence of the approximation is to accumulate the time delay of a photon as it passes through the lens and attribute it to a point on a plane at the distance of the lens. Images then form at stationary points of this "time delay function" as shown in Figure 4.

Figure 4a shows the time delay function in the absence of a lens. As drawn, the line of sight to the source passes through the origin of the lens plane, but for generality in what follows, we take the physical position of the source where it passes through the plane of the lens to be $\vec{s}$. Were an image to appear at sky position $\vec{i}$ (measured in meters), it would be delayed by an amount $\tau_{geom}(\vec{i})$ that reflects the extra path length from the source to the plane of the lens. We have

$$\tau_{geom}(\vec{i}) = \frac{1}{2c} \frac{|\vec{i} - \vec{s}|^2}{\chi_{tran}^{LS}} \quad , \tag{1}$$

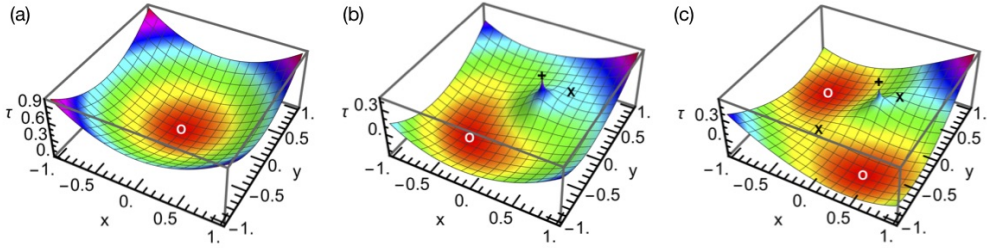

**Fig. 4** Time delay plotted as a function of position in the plane of the lens. Minima of the time of flight are marked with a ○; saddlepoints by a ×; maxima by a +. Panel (a) shows the time delay in the absence of a lens. It increases quadratically outward from the line of sight, which passes through the origin. Panel (b) shows the effect of a circularly symmetric galaxy offset from the line of sight. Panel (c) shows the effect making that galaxy elliptical. The minimum in panel (b) has been split into two minima and a saddlepoint.

where $\chi_{tran}^{LS}$ (measured in meters) is the lens-to-source distance, measured in a co-moving coordinate system that expands with the universe The distinctions between this distance measure and several alternatives are discussed at length in section 8.2 below. In the absence of a lens, the time delay function is parabolic. For simplicity we ignore until section 8 the introduction of a *second* parabolic contribution to the delay, generated between the image and the observer.

Figure 4b shows the effect of introducing a circularly symmetric galaxy that is offset from the line of sight. The galaxy induces a gravitational time delay, $\tau_{grav}(\vec{i})$, which added to the geometric delay gives

$$\tau_{sum}(\vec{i}) = \frac{|\vec{i} - \vec{s}|^2}{2c\chi_{tran}^{LS}} + \tau_{grav}(\vec{i}) \quad . \tag{2}$$

We set the gradient of $\tau_{sum}(\vec{i})$ to zero and then multiply by $c\chi_{tran}^{LS}$ to get

$$0 = (\vec{i} - \vec{s}) + c\chi_{tran}^{LS} \nabla \tau_{grav}(\vec{i}) \quad . \tag{3}$$

We then solve for the stationary points of the combined delay: $\vec{i}_1, \vec{i}_2$ *et cetera*. There are three of these in Figure 4b – a maximum, a minimum and a saddle point. Dividing the physical displacement in the lens plane of the image from the source, $(\vec{i} - \vec{s})$, by $\chi_{tran}^{LS}$ gives the angular deflection of the image, $\vec{\theta} - \vec{\beta}$, where $\vec{\theta}$ is the angular position of the image and $\vec{\beta}$ is the angular position of the source.

Figure 4c shows the effect of making the interposed galaxy elliptical (as is expected, at some level, for all galaxies). The minimum has been split into two minima with a saddlepoint in between. For any finite offset of the lens from the source, this splitting occurs *only* if the lens exceeds a minimum ellipticity.

If the lensed source were a supernova or some other transient, one would first see an image at the deeper minimum, then a second at the higher minimum, then a third at the lower saddlepoint, then a fourth at the higher saddlepoint and finally an image at the maximum. In Figure 5 we show a dozen examples of quadruply lensed quasars formed by elliptically shaped lenses.

## 5 Coordinates, distances, and angles, in words

The treatment of strong lensing in the preceding sections emphasizes words and drawings. Newcomers to strong lensing may find the plethora of coordinates, distances and angles daunting – more so than any subtlety in the underlying physics.

Before we proceed further, many of the quantities already described need to be given algebraic representations. Yet more quantities, to be introduced in the following sections, are also represented algebraically.

Table 1 is a guide to this exploding collection of symbols.

Strong lensing uses two 2-dimensional coordinate systems with parallel axes, one in the plane of the source, and the other in the plane of the lens. Both planes are perpendicular to the line of sight from the observer to the source. We use $\zeta$ as the coordinate along this line of sight in the vicinity of the lens plane, so as not to confuse it with redshift $z$.

As described below in section 8, the curvature of spacetime leads to a distinction between distances measured along the line of sight, *longitudinally*, and those measured perpendicular to the line of sight, *transversely*.

## 6 (De-)magnifications and distortions

Lensed images are distorted. Were one to change the source position, $\vec{s}$, in equation (3), it would produce a change in the image position, $\vec{i}$, that differs in both direction and modulus. The matrix quantity $\partial \vec{i}/\partial \vec{s}$ measures that distortion and is called the "magnification matrix," $M$. Its inverse, $M^{-1}$, can be obtained by taking the gradient of (3),

$$0 = I - \frac{\partial \vec{s}}{\partial \vec{i}} + c\chi_{tran}^{LS} \nabla \nabla \tau_{grav}(\vec{i}) \quad , \tag{4}$$

where the last term on the right is the Hessian of $\tau_{grav}(\vec{i})$ and $I$ is the identity matrix. The second term on the right of the equation is the inverse magnification matrix, $M^{-1}$, yielding

$$M^{-1} \equiv \frac{\partial \vec{s}}{\partial \vec{i}} = \begin{pmatrix} 1 + c\chi_{tran}^{LS} \frac{\partial^2 \tau_{grav}(\vec{i})}{\partial x^2} & + c\chi_{tran}^{LS} \frac{\partial^2 \tau_{grav}(\vec{i})}{\partial x \partial y} \\ + c\chi_{tran}^{LS} \frac{\partial^2 \tau_{grav}(\vec{i})}{\partial x \partial y} & 1 + c\chi_{tran}^{LS} \frac{\partial^2 \tau_{grav}(\vec{i})}{\partial y^2} \end{pmatrix} \quad . \tag{5}$$

The magnification matrix and its inverse are symmetric for well behaved gravitational time delays, and can be diagonalized. If, after diagonalization, the diagonal terms of $M^{-1}$ are both positive (or both negative), the image is a minimum (or maximum) of the time of flight and have even parity. If the diagonal terms have opposite signs, the image is a saddle point and has the opposite handedness of the source.

For strongly centrally concentrated lenses (as is the case for most galaxies), the second derivatives at the center are large and negative, giving maxima that are highly (and sometimes infinitely) demagnified. High curvature of the time delay surface gives high demagnification. In talking about observations of strong lenses we often ignore these elusive maxima and talk about double and quadruple systems, though in principle the systems in Figures 4b and 4c are respectively, triple and quintuple.

**Table 1** Coordinates, distances, and angles, in words

| symbol | ref | definition |
|---|---|---|
| $S$ | Fig. 1 | source |
| $O$ | Fig. 1 | observer |
| $x$ | Fig. 1 | physical distance from the origin in lens plane |
| $y$ | Fig. 1 | physical distance from the origin in lens plane |
| $\zeta$ | Fig. 1 | physical distance along line of sight from lens plane |
| $z$ | §8 | redshift; not to be confused with $\zeta$ |
| $\beta$ | Fig. 2 | scalar angular displacement of source from lens |
| $\vec{\beta}$ | §4 | angular vector displacement of source from lens |
| $\vec{\theta}$ | §4 | angular vector displacement of image from lens |
| $\theta_E$ | eqn (25) | the Einstein radius of a lens |
| $+$ & $-$ | Fig. 2 | parities of images for minima and saddle points |
| $b$ | Fig. 2 | angular deflection caused by an isothermal spherical lens |
| $\vec{i}$ | §4 | a physical vector position in the image plane |
| $\vec{s}$ | §4 | the physical vector position in the image plane where the line of sight to the source passes through it |
| $T^{LO}$ | §8.3 | lens-to-observer time of flight when no lens present |
| $\tau$ | §4 | *time delay* ; extra time of flight w.r.t. no lens present |
| $\tau_{geom}$ | eqn (1) | the *extra* time of flight due to increased path length |
| $\tau_{grav}$ | §4 | the *extra* time of flight due to gravity |
| $\tau_{sum}$ | §4 | the combined *extra* time of flight |
| $\tau_{shapiro}$ | eqn (11) | same as $\tau_{grav}$ ; calculated from gravitational potential |
| $\chi_{long}$ | §8.2 | a co-moving distance along line of sight |
| $f_K(\chi_{long})$ | §8.2 | a co-moving distance $\perp$ to line of sight |
| $M$ | eqn (5) | magnification matrix: $M^{-1}$ defined as $\frac{\partial \vec{s}}{\partial \vec{i}}$ |
| SIEP | §7 | **S**ingular **I**sothermal **E**lliptical (gravitational) **P**otential |
| SIS+XT | §7 | **S**ingular **I**sothermal **S**pherical potential + e**X**ternal **T**ide |
| $\Phi(\vec{i},\zeta)$ | §9 | Newtonian gravitational potential of lens |
| $\Phi_{2D}(\vec{i})$ | eqn (10) | projection of $\Phi(\vec{i},\zeta)$ onto lens plane |
| $\vec{L}$ | §10.1 | angular position vector of lens on plane of sky |
| $\vec{I}$ | §10.1 | angular position vector on the plane of the sky of a hypothesized lensed image |
| $\Psi(\vec{\theta})$ | eqn (13) | Newtonian potential with lens at origin |
| $\Gamma$ | eqn (26) | the strength of the tidal shear flattening the lens potential |
| $\eta$ | §11 | the "semi-ellipticity" of an elliptical *potential* |
| $q$ | §11 | the minor-to-major axis ratio for an ellipse |

In Figure 4c, where three of the images are arranged more-or-less tangentially to the lens, the corresponding component of the inverse magnification matrix is close to zero and the images are highly magnified in the tangential direction.

A useful mnemonic is that there are three "D's" associated with gravitational lensing: *delay, deflection and distortion*, which depend, respectively, on the zeroth, first, and second derivatives of the time delay function.

The magnification matrix might more properly be called the distortion matrix, but we bow to convention on this. The determinant of the magnification matrix gives the scalar factor by which the areas of sources are multiplied, thereby increasing or decreasing the flux received from the source.

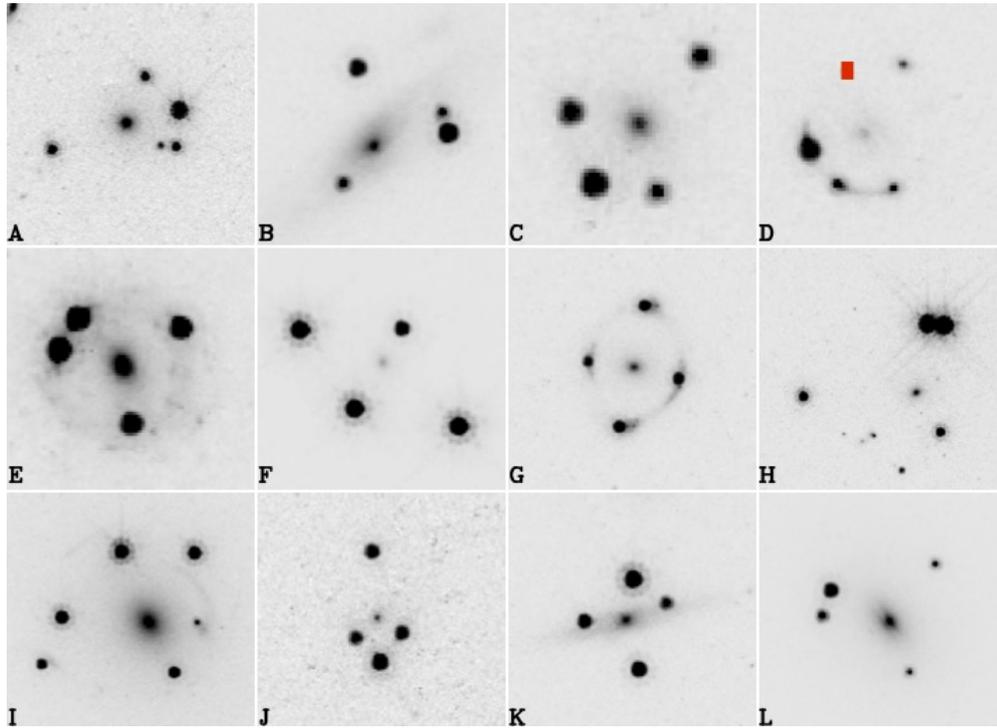

**Fig. 5** Negative images of HST exposures of a dozen arbitrarily selected quadruply lensed quasars. The lensing galaxy is always visible at the center of the field, and in most cases is more extended than the four quasar images. There are faint background galaxies in panels A and H. The two faint starlike images in panel I are images of a *second* quasar. A faint charged particle detection in panel D is masked in red. In panels D, E and G, the galaxy that hosts the quasar can be seen forming something of an "Einstein ring".

# 7 Close pairs, critical curves, and caustics

## 7.1 Close pairs of bright images and critical curves

In Figure 6 we show a slice through the time delay surface along the line connecting a close pair of images. This slice does not tell us which of the two images might be a saddlepoint and whether the other is a minimum or a maximum. A perpendicular slice might show either a maximum or a minimum. But in either case, the maxima and minima within the slice shown are very shallow.

Displacing the source by a small amount or adding mass to the system would produce a correction to the slice through the time delay surface that is linear to first order. If the correction is large enough, we can eliminate these two stationary points. Just prior to their elimination, the two images get arbitrarily close and the curvature along the line connecting them gets arbitrarily small. As magnification is inversely proportional to curvature, that implies infinite magnification along the line connecting the images.

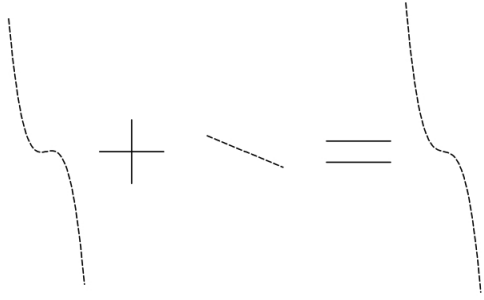

**Fig. 6** A section through the time delay surface for a very close pair of images shows very little curvature, and therefore high magnification. Adding a small gradient to the time delay surface eliminates the two stationary points, and hence the two images. But they go out in a blaze of glory, becoming bright before they disappear.

A locus of points along which a pair of images merge is called a "critical curve". One solves for the locus by setting the determinant of the inverse magnification matrix to zero, $|M^{-1}| = 0$. Pairs of images merging at a critical curve are called a "fold" configuration, and are discussed in exhaustive detail by Gaudi and Petters (2002). The pair of images is part of a larger multiplet – most often a quartet or triplet – the other images of which can be far from the close pair and less highly magnified. The high magnifications of images close to critical curves draws our attention and result in selection effects that favor systems with close pairs over those with more widely spaced image configurations.

The derivative of the zero eigenvalue of $M^{-1}$ along its eigendirection sets a scale over which the critical curve has significant effect – a kind of "swath of influence". The magnification of an image varies inversely as its distance from the critical curve, measured in units of this scale length.

In Figure 7 we show two merging images of a circular source. Both images are lopsided – teardrop shaped – and merge along the peanut-shaped critical curve. While images across folds are, to first order, mirror images of each other, that approximation breaks down if the image is not much smaller than the curve's swath of influence.

## 7.2 Caustics

Critical curves in the image plane map into closed curves in the source plane called "caustics." A source inside a caustic produces two more images than it would outside the caustic." Segments of caustics with finite curvature are called "folds" (Gaudi and Petters 2002). The caustic shown in the inset to Figure 7, an example of a "diamond caustic" has been calculated for the particular critical curve illustrated, a **S**ingular **I**sothermal **E**lliptical **P**otential (SIEP). The diamond caustic is symmetric about the elongation of the 2 dimensional gravitational potential. Its longer axis is parallel to the shorter axis of the elliptical potential. The diamond caustic is described mathematically by the equation for a stretched astroid. The caustic for a **S**ingular **I**sothermal **S**phere with E**x**ternal tide – SIS+XT – is also a stretched astroid.

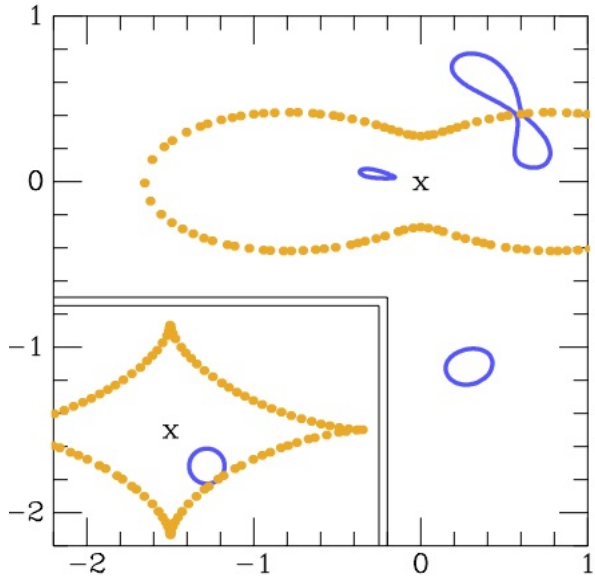

**Fig. 7** A pair of merging images (upper right) of a circular source (inset) lensed by a **S**ingular **I**sothermal **E**lliptical **P**otential (SIEP). An $x$ marks the center of the potential. The peanut shaped (dotted orange) critical curve shows the locus of points at which images can merge, and along which they have infinite magnification. Sources interior to the (orange dotted) diamond caustic (inset) produce four images. A short arc of the circular source lies outside the caustic and produces only two images.

Sources that lie near one of the two symmetry axes of a diamond caustic and close to one of its four "cusps" produce a characteristic triplet of images in a short, gentle arc. Cusp configurations are illustrated in section 13.4 below.

## 8 Distances to and from lenses and sources

For simplicity we assume that we live in a $\Lambda CDM$ universe – one with ordinary matter, "cold dark matter," and a "cosmological constant," $\Lambda$. Distances between points are calculated using the Friedman-Robertson-Walker (henceforth FRW) metric of general relativity.

The FRW metric has two salient features that affect strong lensing. First, the spatial separation between points has an overall scale factor, $a(t)$, that, in our universe, increases monotonically as time progresses. The redshift $z_s$ of observed spectral lines in a distant source tell us that this scale factor was smaller when the light was emitted by a factor $1/(1 + z_s)$ than it is now. Some of the complications introduced by this expansion are described in section 8.2 below.

The second salient feature of the FRW metric is that distances radiating outward from the pole of the (arbitrarily) adopted coordinate system are calculated differently than those calculated perpendicular to those rays. This is a consequence of the curvature of spacetime, which need not be zero, but is very nearly so (Efstathiou and Gratton 2020).

### 8.1 Longitudinal and transverse distances (on the Earth)

A useful analogy is to imagine rays emanating from two sources at the same latitude $\mathcal{B}$ (expressed in radians) on the Earth's surface that are somehow constrained to propagate along its surface. An observer at the Earth's north pole will see them arriving along lines of constant longitude $\mathcal{L}$ separated by some angle $\theta$ (also in radians). The rays arriving at the pole from these sources have traveled a *longitudinal distance* $D_{long}$ proportional to the product of that latitude and the radius of the Earth,

$$D_{long} = (\mathcal{B} - \pi/2) R_\oplus \quad . \tag{6}$$

However, the same pair of sources at a common latitude have a *transverse* separation between them, $D_{tran}$, calculated by multiplying the angular separation by a different factor,

$$D_{tran} = \cos \mathcal{B} R_\oplus \theta \quad . \tag{7}$$

This need for both distances is the direct consequence of the curvature of the Earth's surface.

## 8.2 Co-moving distances

The FRW metric is a recipe for computing distances in a coordinate system that is expanding with the universe. By convention, we use a preferred epoch (today) and origin (ourselves). We call these "co-moving coordinates." Distances that are computed in this coordinate system are called "co-moving distances."

We define the "longitudinal co-moving distance" from a source $S$ to ourselves, $\chi_{long}^{OS}$ in this preferred coordinate system. As FRW spacetimes are generically curved, there is also a "transverse co-moving distance", denoted $f_K(\chi_{long}^{OS})$, where $K = 0, +1,$ and $-1$ for flat, bowl-like and saddle-like topologies, respectively. It is used to calculate co-moving distances perpendicular to the line of sight.

Both co-moving distances are proportional to $a(t)$. For the special case of a flat universe, we have $f_K(\chi_{long}^{AB}) = \chi_{long}^{AB}$, but for positive and negative curvatures the expression is more complicated, involving, respectively, conventional and hyperbolic sine functions.

Transverse and longitudinal co-moving distances both have the special property that $\chi^{AC} = \chi^{AB} + \chi^{BC}$, which is *not* the case for some alternative kinds of cosmological distance.

## 8.3 Time of flight

The time of flight from a source to an observer, $T^{OS}$, is calculated by taking small increments of the longitudinal co-moving distance, $d\chi$, dividing this by the speed of light to get an increment of time, and dividing each increment of time by the FRW scale factor $a(t)$.

We let $d\zeta$ be an increment of physical distance (measured with a meter stick) along the trajectory. Integrating from the source to the observer gives the time of flight. The scale factor is included to account for special relativistic time dilation, making the time of flight longer than it would otherwise be:

$$T^{LO} = \int_{lens}^{observer} \frac{d\chi_{long}^{\zeta}}{d\zeta} (1 + z_\zeta) \frac{d\zeta}{c} \quad \text{at} \ z_{obs} = 0 \quad . \tag{8}$$

Dividing the time-of-flight by the speed of light and integrating we find

$$T^{LO} = \frac{\chi^{OL}_{long}}{c} \qquad (9)$$

where $\chi^{OL}_{long}$ is the longitudinal co-moving distance.

Hogg (1999) gives a clear, pedagogical roadmap to the different kinds of cosmological distances and the time of flight. We use the same words he does to describe them, but adopt different notation. Hogg illustrates their dependences upon redshift, the cosmological constant $\Lambda$ and a non-zero curvature, $\Omega_k \neq 0$. His clarity notwithstanding, the expressions he presents are dauntingly non-trivial, even to experienced astronomers.

## 9 Shapiro gravitational time delay (in the lens plane)

The gravitational contribution to an observed time delay is intrinsically general relativistic in nature. But for sources strongly lensed by galaxies and clusters of galaxies, it can be calculated using the Newtonian gravitational potential, $\Phi(\vec{i}, \zeta)$, where $\vec{i}$ is physical position in the lens plane and $\zeta$ is physical position perpendicular to it. Since the gravitational delay occurs in or close to the lens, it can be calculated as though it occurs instantaneously as the ray traverses the lens plane, at redshift $z_L$. This is the driving virtue of the thin lens approximation of section 4.

A great deal of lens modeling is carried out using what is somewhat misleadingly called a two dimensional "potential" $\Phi_{2D}(\vec{i})$, obtained by projecting the actual potential, $\Phi(\vec{i}, \zeta)$ onto the lens plane. We define it here as

$$\Phi_{2D}(\vec{i}) \equiv \frac{1}{f_K(\chi^{OL}_{long})} \int_{source}^{observer} \frac{2\Phi(\vec{i}, \zeta)}{c^2} d\zeta \quad , \qquad (10)$$

where we have the used the transverse co-moving distance, $f_K(\chi^{OL}_{long})$ to render the projected potential dimensionless, a choice that simplifies calculation in what follows

Following Binney and Merrifield (1998), we derive a general expression for this "Shapiro delay" $\tau_{shapiro}$, by integrating with respect to distance $\zeta$ perpendicular to the lens plane,

$$\tau_{shapiro} = \int_{source}^{observer} \frac{2\Phi(\vec{i}, \zeta)}{c^2} \frac{d\zeta}{c} \quad , \qquad (11)$$

where the use of perpendicular distance $\zeta$ is appropriate for small deflection angles. The Shapiro delay is then given by

$$\tau_{shapiro} = \frac{f_K(\chi^{OL}_{long})}{c} \Phi_{2D}(\vec{i}) \quad . \qquad (12)$$

In the next section, we solve for the *angular* position of an image on the sky, $\vec{I}$, rather than the corresponding spatial position $\vec{i}$ within the lens plane. To avoid

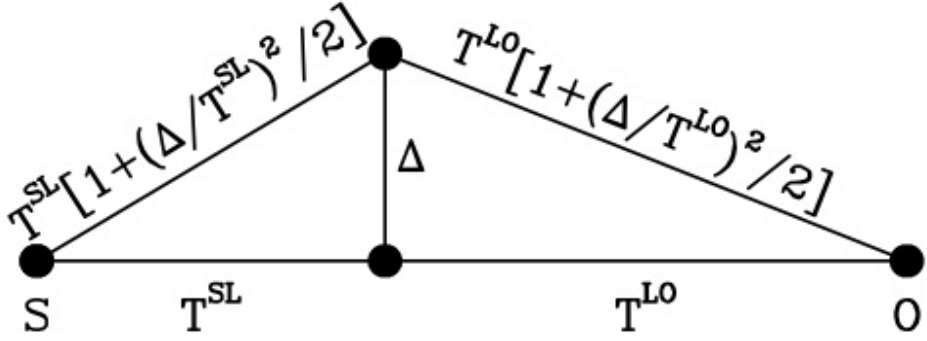

**Fig. 8** Times of flight between source and lens, $T^{SL}$, and between lens and observer, $T^{LO}$, for a lens directly on the line of sight from the observer to the source. For images that are close to the lens, the Pythagorean theorem and the small angle approximation ($\Delta \ll T^{SL}$ and $T^{LO}$) give the times of flight from the source to the image, $T^{SL}\left[1+(\Delta/T^{SL})^2/2\right]$, and the time of flight from the image to the observer, $T^{LO}\left[1+(\Delta/T^{LO})^2/2\right]$.

confusion, we represent the 2D potential as a function of angular position,

$$\Psi_{2D}(\vec{I}) = f_K(\chi_{long}^{OL})\Phi_{2D}\left(\frac{f_K(\chi_{long}^{OL})}{1+z_L}\vec{I}\right) \quad , \tag{13}$$

where $f_K(\chi_{long}^{OL})$, the transverse co-moving distance, turns the angle $\vec{I}$ into a co-moving distance and the factor $1+z_L$ accounts for the Hubble expansion.

## 10 Two geometric delays

### 10.1 Delays as observed in the lens plane

The total time delay $\tau_{sum}$ is the sum of three terms: the general relativistic Shapiro delay, and *two* geometric terms, one from the lens plane to the observer and the other from the source to the lens plane. These must be calculated for an observer at rest with the lens plane for consistency with the gravitational delay.

The transverse co-moving distance, $f_K(\chi_{long}^{OL})$ was defined in section 8.2 to transform the angular separation between points perpendicular to the line of sight into co-moving distances. For the lens system illustrated in Figure 8, it converts the angle between the lens and the image, into $\chi^{LI}$:

$$\chi^{LI} = |\vec{L}-\vec{I}|f_K(\chi_{long}^{OL}) \quad , \tag{14}$$

where $\vec{L}$ and $\vec{I}$ are angular positions on the sky.

We take our lens to be thin and our deflection angles to be small, so that the lens and the images of the source can be taken to lie in a plane at the same redshift.

As noted in section 8.2, the physical distance between points frozen into the Hubble expansion measured at redshift $z_2$ is larger (or smaller) than that measured at $z_1$ by the factor $(1+z_2)/(1+z_1)$. Hence the physical distance $\Delta^{LI}$ measured between the lens and an image in the same plane at redshift $z_L$ is smaller than the co-moving distance that would be measured today, $\chi^{LI}$, by a factor $1/(1+z_L)$:

$$\Delta^{LI} = \frac{1}{1+z_L}\chi^{LI} \quad , \tag{15}$$

and the time of flight measured in the lens plane is

$$T^{LI} = \frac{1}{c}\frac{1}{1+z_L}\chi^{LI} \quad . \tag{16}$$

Using the small angle approximation and the Pythagorean Theorem we get the total times of flight from the image to the observer and from the source to the image:

$$T^{IO} \approx T^{LO}\left[1 + \frac{1}{2}\left(\frac{T^{LI}}{T^{LO}}\right)^2\right] \quad \text{and} \quad T^{SI} \approx T^{SL}\left[1 + \frac{1}{2}\left(\frac{T^{LI}}{T^{SL}}\right)^2\right] \quad . \tag{17}$$

The time delay $\tau$ was defined to be the *extra* time of flight compared to the unlensed time of flight. Hence the geometric contribution to the time delay is

$$\tau^{SO}_{geom}(\vec{I}) = T^{LO}\frac{1}{2}\left(\frac{T^{LI}}{T^{LO}}\right)^2 + T^{SL}\frac{1}{2}\left(\frac{T^{LI}}{T^{SL}}\right)^2 \quad \text{or} \tag{18}$$

$$\tau^{SO}_{geom}(\vec{I}) = \frac{1}{2}\left[\frac{T^{LO}+T^{SL}}{T^{LO}T^{SL}}\right]\left(T^{LI}\right)^2 \quad . \tag{19}$$

## 10.2 The "effective" time of flight

The square bracketted term in equation (19) will remind readers of the effective electrical resistance of two resistors connected in parallel in a circuit. By analogy, we define an "effective time of flight"

$$T^{SO}_{eff} \equiv \left[\frac{1}{T^{LO}} + \frac{1}{T^{SL}}\right]^{-1} \quad . \tag{20}$$

The frequency with which this effective quantity appears in strong lensing calculations warrants its being give a special designation, despite its not being the time of flight to anything. We tabulate values of $T^{SO}_{eff}$ in section 10.5 below. The redshifts typical of strongly lensed quasars and the galaxies lensing them are $z_S = 2.0$ and $z_L = 0.5$, giving a $T^{typical}_{eff} \sim \mathcal{O}(1/7)$ the age of the universe.

## 10.3 Net time delay

Summing the gravitational and geometric contributions to the time delay (as measured by an observer in the lens plane) given by equations (12) and (19) we have

$$\tau_{sum}^{SO}(\vec{I}) = \frac{\chi_{long}^{OL}}{c}\Psi_{2D}(\vec{I}) + \frac{1}{2}\left[\frac{1}{T_{eff}^{OS}}\right]\left(T^{LI}\right)^2 \quad \text{at } z_{lens} \quad . \tag{21}$$

## 10.4 Images are seen at stationary points of the time delay

As discussed in sections 3 and 7, images of the lensed source appear at stationary points of the time delay, where $\nabla\tau_{sum}^{SO}(\vec{I}) = 0$. The gravitational and geometric contributions to the gradient are equal and opposite, and cancel at minima, maxima and saddlepoints.

We have
$$0 = \frac{f_K(\chi_{long}^{OL})}{c}\nabla\Psi_{2D}(\vec{I}) + \frac{1}{2}\left[\frac{1}{T_{eff}^{OS}}\right]\nabla\left(T^{LI}\right)^2 \quad . \tag{22}$$

Substituting the expression for the time of flight $T^{LI}$ in equation (16) and for the angular separation $\vec{I} - \vec{L}$ in equation (14) gives

$$0 = \frac{f_K(\chi_{long}^{OL})}{c}\nabla\Psi_{2D}(\vec{I}) + \frac{1}{2}\left[\frac{1}{T_{eff}^{OS}}\right]\nabla\left(\frac{1}{c}\frac{1}{1+z_L}|\vec{I}-\vec{L}|f_K(\chi_{long}^{OL})\right)^2 \quad . \tag{23}$$

Noting that $\nabla|\vec{I} - \vec{L}|$ is a unit vector in the direction $\vec{I} - \vec{L}$ yields

$$0 = \frac{f_K(\chi_{long}^{OL})}{c}\nabla\Psi_{2D}(\vec{I}) + \left[\frac{1}{T_{eff}^{OS}}\right]\left(\frac{1}{c}\frac{1}{1+z_L}f_K(\chi_{long}^{OL})\right)^2(\vec{I}-\vec{L}) \quad . \tag{24}$$

Equation (24) is a variant of what is called "The Lens Equation." It gives the angle between an image and the center of the potential, $\vec{I} - \vec{L}$, in terms of the gradient of an appropriately defined two dimensional gravitational potential.

Most pedagogical treatments of strong lensing take the lens equation as the starting point for developing subsequent results. It involves quantities called "angular diameter distances" that we have assiduously avoided mentioning. Instead of these, our variant of the lens equation involves times of flight, which arise as a natural consequence of beginning with Fermat's principle.

Our hope is that newcomers to the field find it easier to think in terms of time of flight, letting them postpone dealing with the conventional approach until such time as they must communicate their results in presentations.

## 10.5 Effective time of flight tabulated

In Table 10.5 of the effective time of flight for a flat, $\Lambda = 0.7$ cosmology for some representative lens and source redshifts. Linear interpolation on the tabulated values gives a good first approximation. One must, however, return to the original equations if one's ultimate goal is precision cosmology.

**Table 2** Effective time of flight $T_{eff}$ for respresentative source and lens redshifts

| $z_L$ | 0.1 | 0.3 | 0.5 | 0.7 | 1.0 |
|---|---|---|---|---|---|
| $z_S$ | | | | | |
| 0.2 | 0.041 | | | | |
| 0.5 | 0.064 | 0.072 | 0.000 | | |
| 1.0 | 0.070 | 0.115 | 0.106 | 0.073 | 0.000 |
| 2.0 | 0.072 | 0.127 | 0.137 | 0.130 | 0.106 |
| 5.0 | 0.068 | 0.117 | 0.129 | 0.129 | 0.120 |
| 9.9 | 0.061 | 0.097 | 0.104 | 0.104 | 0.098 |

# 11 Pointlike and Extended Sources: QSOs, SNe and galaxies

Of order $\sim 350$ quasars have been observed to be strongly lensed (Lemon et al. 2023), of which $\sim 60$ are quadruple and the rest are double. The optical emission of quasars arises from accretion disks that are much too small to be resolved by today's telescopes – hence the term quasi-stellar objects – so QSOs are effectively point sources. Supernovae are likewise effectively point sources. In both cases the observables are the positions and fluxes of point images.

By contrast the galaxies that host QSOs and SNe can often be spatially resolved. They make it possible to see the distortions (and magnifications) introduced by strong lenses.

For the sake of simplicity, we consider here only two kinds of sources – pointlike and minimally extended. The latter are described by six numbers: two celestial coordinates, a flux, a semi-major axis, an axis ratio, and an orientation. Sources that are more than minimally extended and sufficiently bright can often be decomposed into pointlike and minimally extended subcomponents.

# 12 An archetypical lens model: the singular isothermal elliptical potential with a parallel external tide

In the same way that our archetypical observed strong lensing system of section 2 is intended as a starting point for thinking about observations, we adopt an archetypical model as a starting point for modelling strong lenses.

In section 2 we modelled our observational archetype with a singular isothermal sphere (SIS), which has the property that images of sources are deflected radially by an angle $\theta_E$.

In section 9 we showed that the gravitational part of the time delay is proportional to a two-dimensional projected gravitational potential given by equation (10). The 2D potential takes a particularly simple form for the SIS,

$$\Psi_{2D}^{SIS} = |\vec{\theta}|\theta_E \quad , \tag{25}$$

where the Einstein radius, $\theta_E$ gives the angular displacement of images from their sources. Note that the potential increases linearly with distance $|\theta|$ from the center of the lens, giving equipotentials that are equally spaced. Should a source lie directly behind the lens, it is stretched into a circular "Einstein Ring" with radius $\theta_E$.

The SIS model has another property that argues for its use: it produces the well known flat circular velocity curves observed in galaxies (Bosma 1978).

It can be generalized to include non-circular potentials. The singular isothermal elliptical potential (SIEP) has equipotentials that are similar concentric ellipses with "semi-ellipticity" $\eta$.[1]

The axis ratios of the equipotentials are $q = (1-\eta)/(1+\eta)$. The SIEP is an SIS potential that has been stretched by a factor $1+\eta$ in one direction and shrunk by a factor $1-\eta$ in the other. As equipotentials are closer along the short axis of the potential, the gradient of the gravitational component of time delay is steeper and minima and stationary points of the total time delay are further from the center of the lens.

Elongated equipotentials are also produced when an SIS lens experiences an e**X**ternal **T**ide from a neighboring galaxy or cluster of galaxies. It is analogous to the production of tides on the Earth's surface and gives a projected potential

$$\Psi_{2D}^{SIS+XT} = |\vec{\theta}|\theta_E - \frac{1}{2}\Gamma(\theta_x^2 - \theta_y^2) \quad . \tag{26}$$

where $\Gamma$ is a measure of the strength of the tide and where we have chosen to put the perturbing mass on the $x$-axis of our coordinate system. Whatever shape a lensed image would have had in the absence of the tide, that image is further stretched by $(1+\Gamma)$ along the $y$-axis and squeezed by $(1-\Gamma)$ along the $x$-axis.

We use the unconventional word "scronching" (Ellenberg 2021) to describe stretching a circle or an ellipse by a factor of $(1+Q)$ in one direction and squeezing it by $(1-Q)$ in the other direction, where $Q < 1$. We reserve the word "shear" to describe the *net* deformation that results from multiple scronchings by different components of the lensing potential.

Both sources of flattening may contribute simultaneously. Calculation of the combined potential is greatly simplified if one takes them to be parallel to each other, giving the **S**ingular **I**sothermal **E**lliptical **P**otential plus parallel e**X**ternal **T**ide, abbreviated SIEP+XT$_\parallel$ lens (Luhtaru et al. 2021). We postpone discussion of the unusual, geometric perspective it permits until section 13.5.

# 13 Configurations and distortions of single sources

The defining characteristic of a strongly lensed system is the presence of multiple images of a single (or multiple) source(s). These multiple images are never exact copies either of the source or of each other. They are stretched, sometimes bent, and often

---

[1] The mass surface density contours for the SIEP are *not* elliptical, but the equipotentials for elliptical mass models are substantially rounder than the mass itself and are very nearly elliptical (Luhtaru et al. 2021)

have the opposite handedness of the source (though unresolvably so for QSOs and SNe).

We use the word "configuration" to describe the positions on the sky of the multiple images in a strongly lensed system. In this section we describe typical image configurations, working up from not-quite-strongly lensed singles through pairs, triplets, quartets and quintets, and note the characteristic distortions associated with the different configurations.

## 13.1 Not-quite-strongly lensed single images

In Figure 2 we used a source with a highly idealized shape to show that an SIS produces neither magnification nor demagnification in the radial direction but can either stretch or shrink a source in the tangential direction. A source that lies just outside $\theta_E$ produces a single image just beyond $2\theta_E$, stretched slightly less than a factor of two in the tangential direction.

This direction-dependent magnification has the effect of transforming an elliptical (minimally) extended source into another elliptical image, with a different semi-major axis, axis ratio and orientation. As the axis ratio will differ by at most a factor of two, a single lensed image is *almost* indistinguishable from an unlensed source. But if the source is more than minimally extended, some curvature will be visible, as described in section 13.2.

## 13.2 Diametrically opposed double configurations

Images form at minima, maxima and saddle points of the time of flight. One cannot add either a minimum or a maximum without also adding a saddle point. Single images must therefore be minima. When two images are observed, there must be a third image, somewhere. As discussed in section 4, systems that appear to be double and quadruple are in fact triple and quintuple, but with a highly (or infinitely) demagnified image at a maxium of the time of flight.

Our archetypical strongly lensed system of Figure 2 has images that are diametrically opposed to each other with a lensing galaxy along that diameter. A source at distance $\beta/\theta_E = 0.63$, gives a positive parity image outside the Einstein ring radius $\theta_E$ that is stretched by a factor of 2.58 in the tangential direction. It also produces a

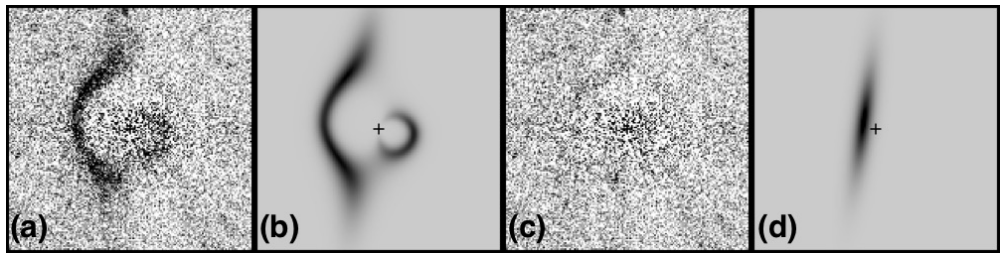

**Fig. 9** (a) An HST image of the galaxy-galaxy lens system SDSS J0912+0029. The lensing galaxy (centered on the +) has been subtracted revealing two images of an edge-on background galaxy. (b) A model of the lensed images; (c) residuals from the model fit; (d) the model of the unlensed source.

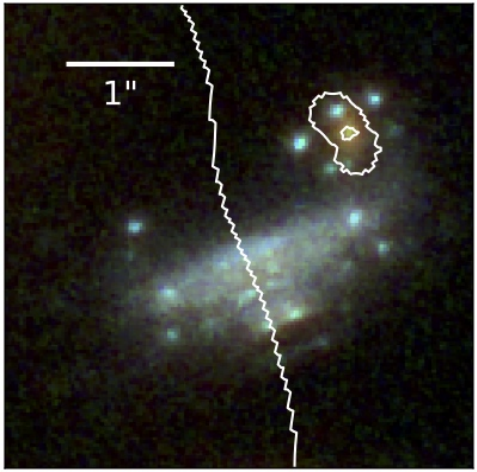

**Fig. 10** Images (e)merging from/into a critical curve (white line) in the cluster SDSS J1226+2152. Dai et al. (2020) estimate the stretching perpendicular to the critical curve to be a factor of $\sim$ 20 at 1" from it, but the bright spots are not elongated, indicating they are unresolved.

second, negative parity image (close to the center of the lens) that is squeezed by a factor of 0.73 in the tangential direction.

Had our source been circular, the outer image would have an axis ratio $q \sim 1/2.58$. But our contrived source is more than minimally extended, so it is curved parallel to the Einstein radius, as as seen in Figure 2. A circular source would give an inner image with $q \sim 0.73$, but with its long axis pointing toward the lens.

Curvature is a generic feature of moderately extended sources, and can be quite dramatic for highly magnified sources. It gives rise to the "arcs" produced in clusters of galaxies that lens background galaxies.

Figure 9 (Bolton et al. 2008) shows the lensed images of a background galaxy that extends well beyond the Einstein ring, $\theta_E$, of the lensing galaxy. Panel a) shows HST data obtained with the F814W filter. Panel b) shows the predictions from a model for the system. Panel c) shows the residual difference between the observations and predictions and panel d) shows the model for the unlensed background galaxy. This image is unlike anything seen in unlensed galaxies, and its interpretation as a strongly lensed system is ironclad.

In Figure 10 we see a galaxy lensed by the cluster of galaxies SDSS J1226+2152 (Dai et al. 2020) stretched perpendicular to a critical curve derived from a model. Pairs of blue knots, presumably mirror images of each other, straddle the critical curve. They must be unresolved because they are not elongated parallel to their separation.

### 13.3 Triplets from straddled saddles

There is a useful distinction to be made between triplets of lensed images formed between two lensing galaxies, and triplets resulting from "naked cusps," whose origin is described in section 13.4 below. We call the former "straddled saddles" because they have two images that straddle a saddlepoint of the 2D gravitational potential. The inner image is a saddlepoint of the light of the time of flight, while the outer two are minima.

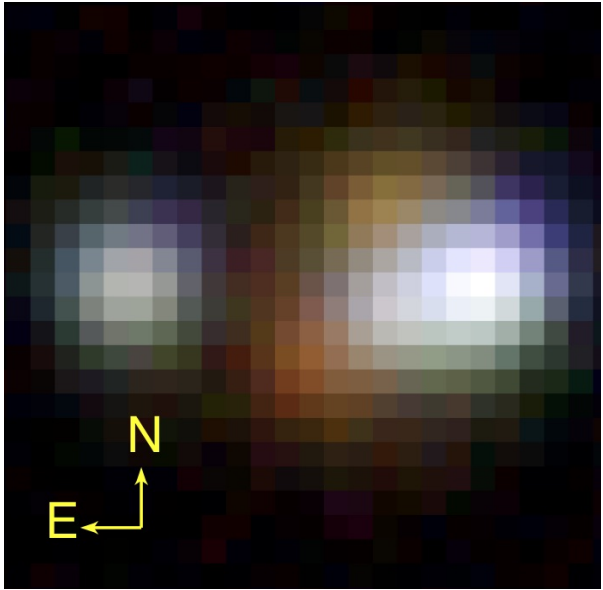

**Fig. 11** WGD0316-41, an example of a "straddled saddle," with three bluish images of the quasar running East-West, passing between two reddish lensing galaxies. The faintest of the three images is a saddlepoint of the time of flight, and lies close to one of the two minima. The galaxies are separated by $\sim 1.4''$.

WGD0316-41 (Agnello, private communication), shown in Figure 11, is an example of a straddled saddle. The middle image does not lie precisely between the two lensing galaxies because the source is offset from the saddlepoint of the 2D potential. Such configurations occur more frequently at the centers of clusters of galaxies (which sometimes have two dark matter concentrations) than between pairs of galaxies.

## 13.4 Triplets associated with cusps

A source that lies along one of the axes of symmetry of a diamond caustic produces a triplet of images that is symmetric around the corresponding axis in the image plane. Mathematically, the diamond caustic is a "scronched" astroid (Falor and Schechter 2022)[2] with four "cusps."

In the same way that the two images near a fold are highly magnified by the same factor (though with opposite parities), the two outer images of a cusp triple are highly magnified and have the same parity. The inner image is twice as highly magnified, and has the opposite parity. Cusps may have either positive or negative net parity.

Starting with a source on one of the symmetry axes and moving perpependicular to it, the cusp relation begins to fail, with two of the images converging and brightening and the other getting fainter. Few, if any investigators explain at what point they chose to call a particular triplet a cusp rather than a fold and a third image.

For most cusp configurations, there is a single diametrically opposed image on the other side of the galaxy whose parity differs from the combined parities of the 3 cusp images.

---

[2] A true astroid is the given by the locus of solutions of the equation $x^{\frac{2}{3}} + y^{\frac{2}{3}} = 1$

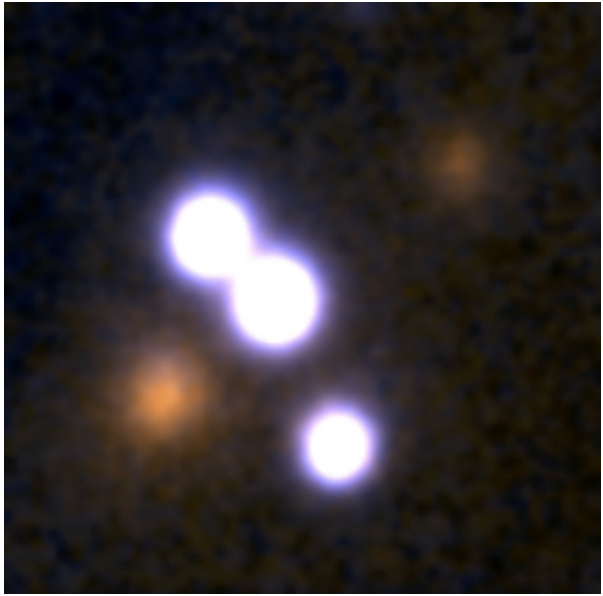

**Fig. 12** J0457-78, which seems to be a naked cusp. There is no hint of a fourth image close to the lensing galaxy. A second, fainter galaxy may be seen to the upper right, in which case the potential may be elongated due to a tide rather the flattening of the lensing galaxy. If a tide is indeed the source of elongation, the topology of the lens is the same as for a straddeled saddle, and the distinction becomes one of degree.

If one starts with the source at the center of the diamond caustic, the four images form a rhombus. As one moves the source along the longer axis of the diamond caustic, three images begin to converge and get further from the center of the lens. The diametrically opposed image (a saddlepoint) approaches the center of the lens. If the ellipticity of an SIEP is greater than 0.5, the fourth image merges with the central, infinitely demagnified image leaving only the three converging images. This configuration is called a "naked cusp." As with the straddled saddle, two positive parity images flank a negative parity image.

Tides can also cause a naked cusp. The system J0457-7820 (Lemon et al. 2023) shown in Figure 12 looks at first like a naked cusp configuration. But there is a second, fainter galaxy in roughly the direction of the elongation of the potential. If this second galaxy were too faint to be observed, the system would indeed be a naked cusp. But as it is visible, the topology is that of a straddled saddle. The tidal interpretation is supported by the absence of strong flattening of the lensing galaxy.

### 13.5 A taxonomy for quadruple image configurarions

Most four-image configurations arise from flattened potentials, either those of galaxies or those of clusters of galaxies. Discounting translations, rotations and scalings, the space of configurations is, at a minimum, 3 dimensional. Schechter (2022) developed a geometric classification scheme for quads that attempts to refine of the qualitative scheme of Saha and Williams (2003).

In this taxonomy, three dimensionless parameters are needed to describe each quad: its non-circularity, its symmetry about one or the other of its diagonals, and the position angle of that diagonal on the sky.

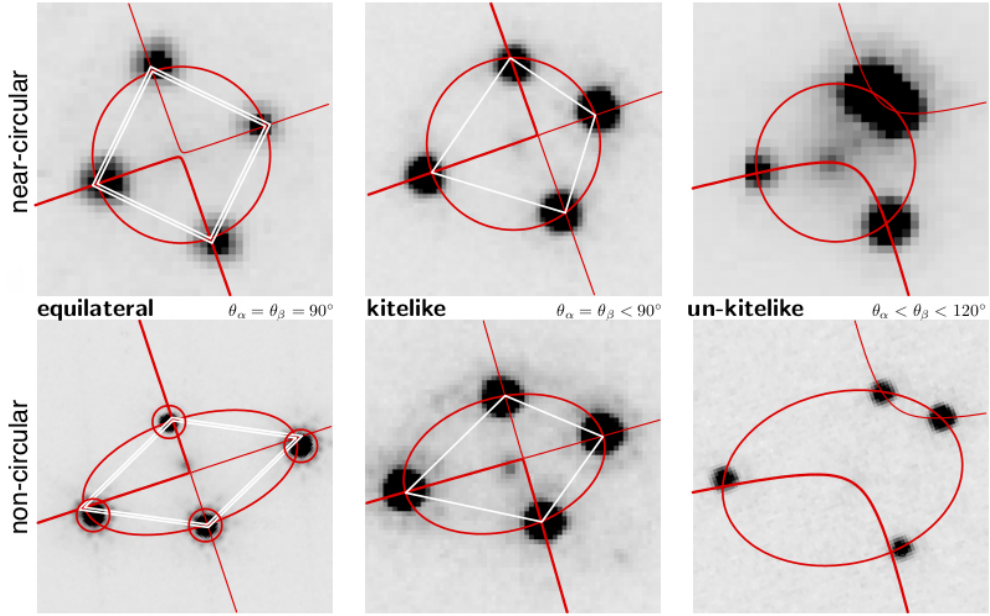

**Fig. 13** A classification scheme for quadruple configurations of lensed quasars, illustrated with HST F814W exposures. The "equilateral", "kitelike" and "un-kitelike" configurations are distinguished by their deviations from axisymmetry along one or the other of the two diagonals. "Non-circular" configurations are stretched versions of "nearly circular" configurations.

The classification scheme is *intentionally* imprecise. If on visual ("eyeball") examination, without using a ruler, the four images might conceivably lie on a circle, the system is "circular." Similarly, if without measuring, the four images might conceivably be symmetric about one *or* the other diagonals, the system is "kitelike." And if the system is conceivably symmetric about both diagonals, the system is "equilateral."

The scheme is illustrated in Figure 13. The bottom row shows the clearly non-circular configurations while the top row shows the nearly circular ones. The leftmost column shows the "equilateral" configurations and the central column shows the "kitelike" ones. The rightmost column shows configurations that are "un-kitelike."

Red curves are superposed on the six lenses shown in Figure 13, with the images lying very close to the four points where the curves intersects. We take the closed curves to be ellipses and the other two curves to be the branches of a hyperbola, which is a geometric representataion of a specific model, the archetypical SIEP+XT$_\parallel$ model, described in section 12.

Other models produce curves that will not, in general, be ellipses or hyperbola. The "lens equation", defined in section 10.4, can be decomposed into a radial component (with respect to the source) and an azimuthal component. Each component of the lens equation gives a locus in the image plane. Images must satisfy both components, and must therefore lie where the loci cross.

To the extent that the loci look *roughly* like hyperbolae, which they do precisely for elliptical potentials (Witt 1996), and ellipses, which they do in general for potentials with a dominant quadrupole component and precisely isothermal elliptical potentials (Wynne and Schechter 2018), the geometric taxonomy is still useful.

The hyperbolae for the symmetric kitelike and equilateral configurations in Figure 13 have very small semi-major axes – so small that they are almost indistinguishable from crosses. Yet for the two un-kitelike configurations, the branch of the hyperbola passing through the close pair of images barely intersects Wynne's ellipse.

The longer branch of the Witt's hyperbola has the theoretical property that it always passes through the center of the lens. Only the positions of the images (and not their fluxes) were used in constructing the hyperbolae shown in Figure 13. The fact that they pass so close to the observed lensing galaxy supports the proposition that the lensing potentials of relatively isolated galaxies are very nearly centered on those galaxies (Schechter and Luhtaru 2024).

The non-circular configurations look like "scronched" (Ellenberg 2021) versions of the corresponding the circular configurations, squeezed along one of the asymptotes of Witt's hyperbola and stretched along the other. These asymptotes are the symmetry axes of the underlying gravitational potential, which is elongated *perpendicular* to Wynne's ellipse (Wynne and Schechter 2018).

The fluxes of the four images are not used in the taxonomy because the effects of micro-lensing can be quite large (Weisenbach et al. 2021). Note though that in the absence of micro-lensing, the four images in the equilateral configurations are expected to be nearly equal in flux.

For the un-kitelike configurations, the fluxes for the two closest images are expected to be roughly equal, and brighter than the faintest image by a factor that depends inversely upon their separation, as described in section 7.1.

Readers are invited to judge for themselves the utility of the geometric taxonomy presented at length in this section by first trying to apply it to Figure 5 and then examining Figure 14, which shows Witt hyperbolae and Wynne ellipses overlaid. Imprecise as the scheme might be, it puts words to otherwise unarticulated geometric impressions that the images give.

### 13.6 Monopole + quadrupole quintets

The systems described at length in the previous section as quartets might be more accurately described as failed quintets. Had their potentials been slightly less cuspy, they would have produced observable fifth images close to the centers of their lenses. Instead those fifth images are highly or infinitely demagnified (Keeton 2003).

Though this is the rule for quasars lensed by galaxies, clusters of galaxies appear to be less cuspy and *do* show such fifth images.

### 13.7 Double lens quintets

At least three lensed quasars have five images, with the fifth image a saddlepoint attributable to a second lensing galaxy. The five brightest sources in Figure 15 are images of the quasar PS J0630-1210 (Ostrovski et al. 2018). The faintest and most central of these is a saddle point of the time of flight, and lies between the two faint red galaxies.

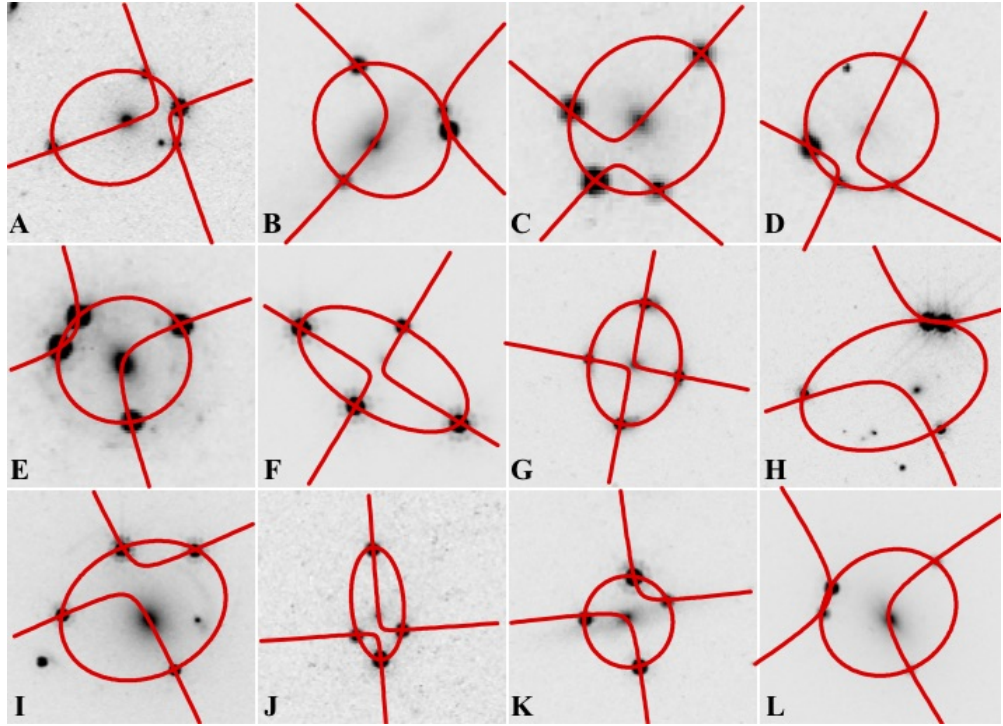

**Fig. 14** The red curves show Witt hyperbolae and Wynne ellipses overlaid on the quadruply lensed quasars of Figure 5. The authors would call systems $D, E, K$ and perhaps $B$ and $L$ circular according to the geometric taxonomy advocated in this section. They would call $A, C, D, I, J$ and $K$ kitelike and $F$ and $G$ equilateral.

### 13.8 Sextets, septets *et cetera*

Starting with a single lens, one can add an arbitrary number of additional images by adding additional lenses. If the added potentials are steeper than isothermal, each additional lens will add, in the simplest instances, a saddlepoint and an infinitely demagnified maximum. If they are less steep than isothermal, they add a saddlepoint and a maximum of finite magnification, provided that the central surface mass density of the added lens exceeds a value that depends upon the redshifts of the source and the lens.

### 13.9 The Witt-Wynne Construction: Geometric not Algebraic

While the archetypical model in introduced in section 2 can be represented by the algebraic expression for it gravitational potential, equation (26) (Luhtaru et al. 2021), we represented it *geometrically* in section 13.5 with the Witt-Wynne construction. One does not need to know *any* physics to use it in modelling stronly lensed systems.

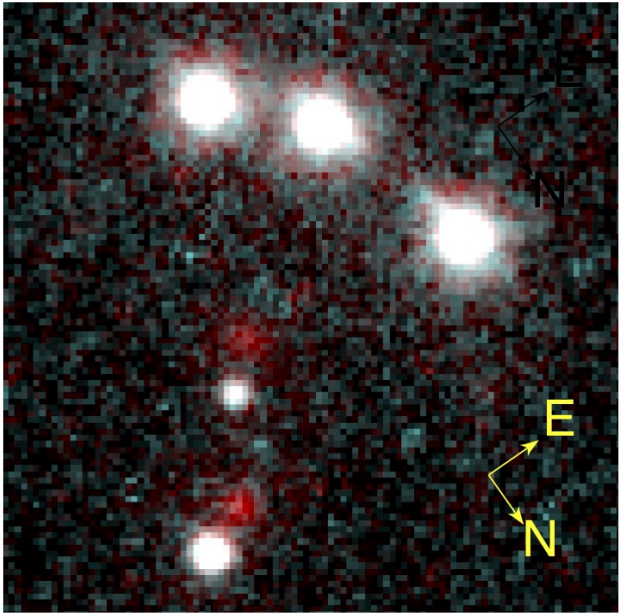

**Fig. 15** A double lens quintet. Two galaxies, faint red blobs in this HST composite, produce five blue images of the quasar J0630-1210. The central, odd parity image lies in the saddle of the galaxies' gravitational potential.

This is analogous to the case of Keplerian ellipses, which are geometric constructs that can be used to model planetary orbits without any understanding of Newton's Universal Law of Gravitation. In both cases one gets a different perspective on the model from the geometric construction than one gets from its algebraic representation. It leads to insights that do not come readily from mathematical expressions.

Readers can gauge for themselves the success of the Witt-Wynne construction in reproducing the positions of quadruply lensed quasars by examining Figure 14. System "H" (J0818-26), for which it works least well, has a bright galaxy close to Witt's hyperbola and several fainter ones just inside or just outside Wynne's ellipse.

The archetypical SIEP+XT$_\parallel$ also produces algebraic expressions for the fluxes of the four images. These are not represented in the figure, as they are strongly influenced micro-lensing (Weisenbach et al. 2021).

Though there will be cases where the SIEP+XT$_\parallel$ model fails to capture interesting aspects of some particular system, it has the substantial advantage of permitting the direct, forward calculation of image positions, with no iteration (Falor and Schechter 2022). Comparison of Falor's method with two inverse schemes for solving the lens equation have given speedup factors in excess of twenty thousand.

## 14 Recapitulation

Three "Ds" capture the principal consequences of strong gravitational lensing: **D**elay, **D**e-flection and **D**istortion. Photons traveling from a source to an observer are **D**elayed by the gravitational potential of an intervening lens. They are **D**eflected by gradients in that gravitational potential, increasing the path length. Fermat's principle dictates that images of the source form on trajectories for which the combined time of flight from these two contributions are a minimum, a maximum or a saddlepoint.

A circular source is **D**istorted by *gradients of the gradient* of the potential, which give an inverse magnification matrix. To first order the image of a circular source is stretched and/or squeezed into an ellipse.

Simple circularly symmetric gravitational potentials that are most negative at their centers (e.g. $-GM/r$) produce three images: a minimum, a maximum and a saddlepoint. More complicated potentials add images in pairs. One of these is a saddlepoint, and the other is either a minimum or maximum. Maxima tend to be highly de-magnified, sometimes infinitely, and often unseen.

The **S**ingular **I**sothermal **E**lliptical **P**otential (SIEP) can be treated as the archetypical lens, manifesting most of the basice elements of strong lensing. Adding a parallel **E**xternal **T**ide gives the Witt-Wynne construction, putting the images of a source at the points of intersection of a hyperbola and an ellipse. This model provides a geometric perspective on strong lensing and a good first approximation for many lensed systems.

The generic challenge in strong lensing is to use the shapes and positions of the observed multiple images to infer simultaneously both the photometric structure of the unlensed source and the structure of gravitational potential, which is often dominated by dark matter.

**Acknowledgements.** We thank Caitlin Millard, Aaron Cohen, Luke Weisenbach and Adam Bolton who created and drafted figures for our use. We thank two reviewers for their guidance in clarifying our presentation.

## Declarations

- Funding
  Some of the observations presented herein were made with the NASA/ESA Hubble Space Telescope and obtained from the Space Telescope Science Institute, which is operated by the Association of Universities for Research in Astronomy, Inc. under NASA contract NAS 5-26555. They were carried out under program HST-GO-15652.
- Conflict of interest/Competing interests
  The authors declare that they have no conflicts of interest that are relevant to this article, nor are they affiliated or involvd with any organization or entity with any financial interest or non-financial interest in the subject matter or materials discussed in this manuscript.